\documentclass[referee]{jfm}
\usepackage{graphicx}
\usepackage{epstopdf, epsfig}

\usepackage{lineno}

\usepackage{newtxtext}
\usepackage{newtxmath}
\usepackage{natbib}
\usepackage{hyperref}
\hypersetup{ 
    colorlinks = true,
    urlcolor   = blue,
    citecolor  = black,
    citecolor = blue
}

\newcommand{\RomanNumeralCaps}[1]
\linenumbers

\usepackage{subfigure}

\usepackage{soul}

\shorttitle{Scaling laws for the sound generation of bio-inspired flapping wings}
\shortauthor{Li Wang, Xueyu Ji, John Young, Hao Liu and Fang-Bao Tian}

\title{Scaling laws for the sound generation of bio-inspired flapping wings during hovering flight}

\author{Li Wang\aff{1},
  Xueyu Ji\aff{1},
  John Young\aff{1},
   Hao Liu\aff{2},
  \and Fang-Bao Tian\aff{1}
  \corresp{\email{fangbao.tian@unsw.edu.au}}}

\affiliation{\aff{1}School of Engineering and Technology, University of New South Wales, Canberra, ACT 2600, Australia \aff{2}Graduate School of Engineering, Chiba University, Chiba 263-8522, Japan}

\begin{document}

\maketitle

\begin{abstract}
Bio-inspired flapping wings have been extensively studied for their remarkable aerodynamic performance. Recently, their noise emission has attracted growing interest, but a careful analysis of scaling laws for their sound generation is missing. This work presents scaling laws for the sound generation of bio-inspired flapping wings during hovering flight based on the quasi-steady flow theory and Ffowcs Williams-Hawkings analogy. Direct numerical simulations considering a range of parameters including the Reynolds number, Mach number and wing kinematics confirm that the proposed scaling laws capture the major physics involved and their predictions agree well with the numerical results. The scaling laws can be used as a powerful tool for engineers in the design of micro-aerial vehicles considering both aerodynamics and acoustics performances simultaneously.
\end{abstract}


\section{Introduction}
Bio-inspired flapping wings, as an important way of aquatic and aerial animals moving through air and water, have been shown to have superior aerodynamic and hydrodynamic performance in comparison to the stationary wings in the low Reynolds number ($O(10^3)$ or lower) regime~\citep[see e.g.,][]{lighthill1975aerodynamic,sane2003aerodynamics,wang2005dissecting,wu2011fish,shyy2013introduction,eldredge2019leading,liu2024vortices}. The superior performance is achieved through several unique mechanisms such as delayed stall of leading edge vortex (LEV)~\citep{ellington1996leading}, wake capture~\citep{dickinson1999wing,bomphrey2017smart}, rapid pitching rotation~\citep{bennett1970insect,dickinson1999wing}, unsteady wing-tip vortex~\citep{shyy2010recent}, synchronised passive pitching deformation~\citep{shyy2010recent,nakata2012aerodynamic,Nakata2012jcp,dai2012dynamic,tian2013force,huang2023power}, and clap and fling~\citep{weis1973quick,lighthill1973weis}. For aerial animals and bio-inspired aerial vehicles, the noise generated by flapping wings is also an important factor measuring the aerodynamic performance, as discussed in a few recent studies~\citep{geng2017effect,wang2019acoustics,wang2020numerical,jaworski2020aeroacoustics,Clark2021review}. The study of the sound generation of bio-inspired flapping wings would shed light on the sound control of bio-inspired aerial vehicles.

One of the most fascinating features of bio-inspired flapping wings is the high lift generation at low Reynolds numbers (e.g. $O(10^2)$ to $O(10^3)$) by using unsteady flow mechanisms~\citep{shyy2010recent,shyy2013introduction,eldredge2019leading,liu2024vortices}. The unsteady flow mechanisms are realised through the complicated kinematics which could be one or the combination of stroke, pitching, deviating, twisting and revolving components~\citep{bomphrey2017smart}. Different components contribute to the aerodynamic force generation in different ways~\citep{shyy2010recent,eldredge2019leading,liu2024vortices}. For example, the stroke motion (also known as the translational motion for 2D cases) creates an intense LEV undergoing a delayed stall to generate high lift which is not possible according to conventional laws of aerodynamics~\citep{ellington1996leading,kweon2010sectional}. The stabilised LEV observed for low Reynolds number insects/birds flight has been further elaborated by using experimental measurements~\citep{poelma2006time} and direct numerical simulations~\citep{liu1998computational,liu2009size}, showing that the attached LEV generates low pressure regions enhancing the lift generation during translation. The pitching (rotational) motion of the wing generates the lift during stroke reversal through the Kramer effect and the wake capture~\citep{dickinson1999wing,sane2002aerodynamic}. The wing revolving/rotating enhances the lift generation by stabilising the LEV through the spanwise pressure gradient, centripetal force, and Coriolis force~\citep{ozen2012three,jardin2015coriolis}. In addition, the dynamic bending of flexible wings delays the breakdown of LEV near the wing-tip augmenting the production of stroke-averaged vertical force~\citep{nakata2012aerodynamic}.

Even though the unsteady kinematics could enhance the force production significantly, the force variations are also larger compared with that of the steady wings. The force variations may have significant consequences on the acoustics generated by the wings especially the Gutin sound~\citep{Clark2021review}, which have not been well explored~\citep{nedunchezian2019effects,wang2020numerical}. The flapping-wing induced sound has drawn the attention from zoologists in very early stages. For example, \citet{ewing1968courtship} found that the complex sound produced by drosophila has a similar frequency to the wing beat frequency during flight. One of the earliest rigorous study on the aeroacoustics of flapping wing can be originated to the experimental study conducted by~\citet{sueur2005sound} who measured the acoustic outputs of flies in tethered flight, finding that the first frequency component is associated with the wing beat frequency dominating the front of the flyer and the second harmonic is dominate at two sides. The two-dimensional numerical study conducted by~\citet{bae2008aerodynamic} identified two different mechanisms of sound generation by flapping wings, where the frequency of the primary dipole tone is the same as the frequency of the transverse wing, while the vortex edge scattering during a tangential motion is responsible for other higher frequency dipole tones. \citet{bae2008aerodynamic} also pointed out that the wing-vortex effects are more prominent in forward flight resulting in a broader high frequency noise band. To further elaborate the sound generation by flapping wing in real world, three dimensional numerical studies have been conducted by \citet{inada2009numerical,geng2017effect,wang2020numerical}. Specifically, the pressure variation on the wing surface is the major source of the acoustics generated by the flapping wing~\citep{inada2009numerical}, which is consistent with the classical acoustics analogy theory. \citet{geng2017effect} adopted a hydrodynamic/acoustic splitting approach to numerically study the sound generated by a forward-flying cicada. By using a prescribed wing kinematics considering the wing deformation, they found that the flexibility is beneficial in the noise reduction in all directions. Inspired by the numerical simulations considering fully flexible flapping wings which demonstrated the enhanced aerodynamic performance by appropriate flexibility, \citet{wang2019acoustics} considered the fluid–structure-acoustics interaction of flapping wings and found that an aerodynamics-dominated flexible wing experiences sound reduction and aerodynamic performance enhancement, while the flexibility of inertia-dominated wings strengths the sound outputs. \citet{nedunchezian2019effects} numerically studied the sound generation by flapping wings with various kinematics with Ffowcs Williams-Hawkings (FW-H) equations and found that the high lift and the low sound are found with the large-amplitude flap and pitch, and the delayed pitching motion. \citet{nedunchezian2019effects} also pointed out that the delayed pitching motion of the wing may aim to decrease the sound generation instead of enhancing the lift force. The three-dimensional fluid-structure-acoustics interaction considering three-dimensional flapping wings in hovering flight by~\citet{wang2020numerical} confirmed that the aerodynamics-dominated wing could experience noise reduction and lift enhancement simultaneously.

Even though scaling laws for the aerodynamics of flapping-wing have been developed~\citep{Dewey2013jfm,kang2013scaling,lee2015scaling,quinn2014,Floryan2017jfm,sum2019scaling,Senturk2019aiaa,ayancik2019}, they have not been considered for the acoustics except for two recent works~\citep{wang2020numerical,Seo2020bb}. Specifically, a simple scaling law for the noise output from flapping wings versus their aerodynamic power requirements was developed in \citet{wang2020numerical}. \citet{Seo2020bb} reported that the sound power efficiency is proportional to $f^2$. The present work is to develop scaling laws for the aeroacoustics of flapping wings during hovering flight based on the quasi-steady flow theory and Ffowcs-Williams-Hawkings analogy. Specially, we first consider purely translating wings in two-dimensional domain. We then consider two-dimensional flapping wings with a combined translating and pitching motion. Finally, we consider three-dimensional flapping wings during hovering flight.
The rest of this work is organised as follows: the physical formation of flapping wing considered in this study is introduced in \S~\ref{sec:phpro}; the numerical method used for the direct numerical simulation of the aerodynamics and aeroacoustics of flapping wing is introduced in \S~\ref{sec:nummthd}; the results and discussions are then presented in \S~\ref{sec:resdic}; and a final conclusion is given in \S~\ref{sec:conc}.

\section{Physical problem}\label{sec:phpro}
\begin{figure}
  \centerline{\includegraphics[width=4.5in]{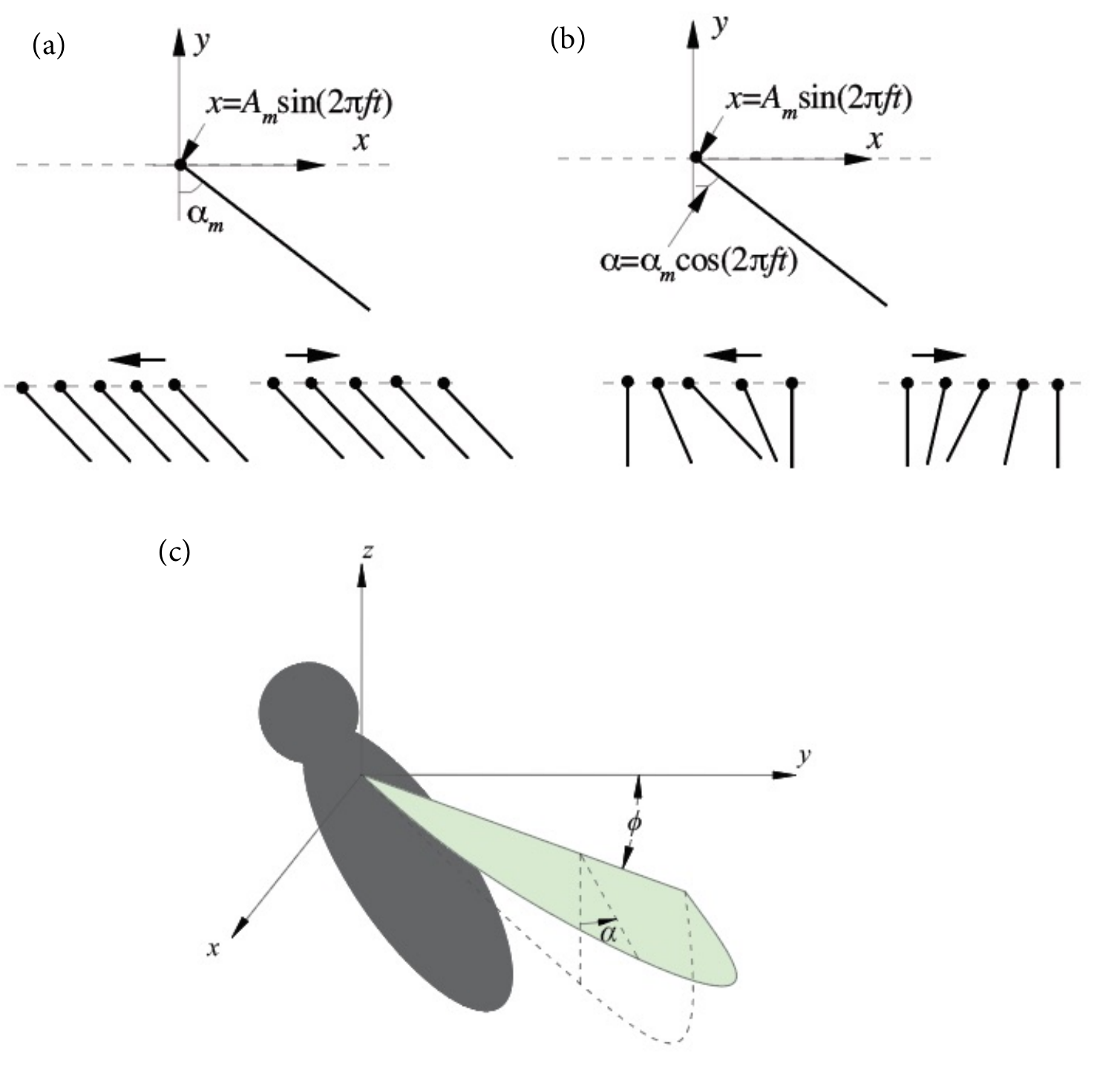}}
  \caption{A schematic of flapping wings: (a) 2D translating wing at a constant angle of attack, (b) 2D translating and pitching wing, and (c) 3D flapping wing in hovering flight.}
\label{fig:sch}
\end{figure}

To develop the scaling laws for hovering flapping wings, the sound generation by flapping wings in both two-dimensional (2D) and three-dimensional (3D) spaces is considered, as shown in Fig.~\ref{fig:sch}. For the simplicity, the wing translating at a constant angle of attack namely $\frac{\pi}{2}-\alpha_m$ in the horizontal direction is first considered, as shown in Fig.~\ref{fig:sch} (a). The kinematics of the translation is prescribed by the following equation
\begin{equation}
    x = A_m \sin(2 \pi f t),
\label{eq:trankint}
\end{equation}
where $A_m$ is the amplitude of the translation, $f$ is the frequency, $x$ is the position of the leading edge and $t$ is the  time. For the flapping wing with a combined translating and pitching motion, as shown in Fig.~\ref{fig:sch} (b), the pitching montion is prescribed as
\begin{equation}
  \alpha = \alpha_m\cos(2 \pi f t),
    \label{eq:pitckint}
\end{equation}
where ${\pi}/{2}-\alpha_m$ is the maximum angle of attack. The dimensionless parameters governing this problems include translating amplitude, $a_m$, Reynolds number and Mach number,
\begin{equation}
    A_m/\bar{c}, \quad a_m, \quad Re = \frac{\rho U \bar{c}}{\mu}, \quad M = \frac{U}{c_s},
    \label{eq:governpara}
\end{equation}
where $U=2 \pi f A_m$, $\bar{c}$ is the mean chord length, $c_s$ is the speed of sound, and $\rho$ and $\mu$ are respectively the density and viscosity of the fluid. In this work, low Reynolds numbers and Mach numbers are considered to limit the scope for insect flight, as shown in Tab.~\ref{tab:para}. Other parameters of kinematics are also selected based on insects. For the 3D flapping wing undergoing stroke and pitching motion, as shown in Fig.~\ref{fig:sch} (c), the kinematics is governed by
\begin{equation}
     \phi = \phi_m \sin(2 \pi f t), \quad \alpha = \alpha_m \cos(2 \pi f t),
    \label{eq:pitckint3d}
\end{equation}
where $\phi_m$ and $\alpha_m$ are respectively the amplitude of stroke and pitching. The flow of the 3D hovering flapping wing is governed by $\phi_m$, $\alpha_m$, Reynolds number and Mach number as defined in Eq.~\ref{eq:governpara} with $U=2 \pi f \phi_m \bar{c} AR$, and the aspect ratio $AR$ of the wing. A range of parameters are examined to construct the scaling law for the sound generation, as shown in  Tab.~\ref{tab:para}. Therefore, the fluid flow is considered as weakly-compressible. 
The lift and drag coefficients are defined as 
\begin{equation}
    C_L = \frac{F_L}{0.5 \rho U^2 A}, \quad C_D = \frac{F_D}{0.5 \rho U^2 A}, 
\end{equation}
where $F_L$ and $F_D$ are respectively the sum of force exerted by the fluid on the wing in the vertical and horizontal directions which can be directly obtained by using the sum of the Lagrangian force introduced in the immersed boundary method (IBM), and $A$ is the area of the wing which is $\bar{c}$ for 2D cases. The sound pressure is defined as $\Delta p = p - \bar{p}$ with $\bar{p}$ being the time averaged pressure, and the time averaging is conducted using ten cycles after the periodic state (cycle-to-cycle variation in the forces is negligible) is achieved. All the sound pressure presented here are scaled by the flow density and sound speed in quiescent state.


\begin{table}
  \begin{center}
  \begin{tabular}{cccccc}
   2D & $M$  & Re & $A_m/\bar{c}$ &  $\alpha_m$ \\[3pt]
      &0.04-0.1 & 200-800   & 0.8-1.4  & $30^o-60^o$ \\
   3D & $M$  & Re & $\phi_m$ & $\alpha_m$ & AR\\[3pt]
      &0.04-0.1 & 200-800   & $30^o-60^o$ & $45^o-60^o$ & 2.0-4.0 \\
  \end{tabular}
  \caption{Parameters for the 2D and 3D flapping wings.}
  \label{tab:para}
  \end{center}
\end{table}

\section{Numerical methods}\label{sec:nummthd}
The unsteady aerodynamics of bio-inspired flapping wings is directly solved by using an immersed boundary--lattice Boltzmann method (IB-LBM), where  the weakly compressible flow is solved by the LBM and the  no-slip boundary condition is implemented by an IBM. Without loss of generality, here we use 2D IB-LBM as an example to introduce the method. In the multi-relaxation-time (MRT)-based LBM, the particle distribution function $g_i$ is given as~\citep{Coveney2002,guo2008analysis,xu2018novel}
\begin{equation}
g_i(\boldsymbol{x}+ \boldsymbol{e}_i \Delta t, t+\Delta t)=g_i(\boldsymbol{x}, t) -\Omega_i(\boldsymbol{x},t)+\Delta t G_i, \quad i=1,2,...9,
\end{equation}
where $\Delta t$ denotes the time step, $\boldsymbol{e}_i$ is the discrete velocity, subscript $i$ indicates  the $i$-th direction,  $\boldsymbol{x}$ is the space coordinator, $\Omega_i$ the collision operator, and $G_i$ is the effect of the body force. For the MRT LBM, $\Omega_i$ and $G_i$ are given as,
\begin{eqnarray}
\Omega_i&=&-(M^{-1}SM)_{ij}[g_j(\boldsymbol{x},t)-g_j^{eq}(\boldsymbol{x},t)],\\
G_i&=&[M^{-1}(I-S/2)M]_{ij}F_j,
\end{eqnarray}
where $M$ is a transform matrix with a dimension of $9\times9$ for the 2D nine discrete velocity (D2Q9) model and $19\times19$ for 3D nineteen discrete velocity (D3Q19), $S$ is the collision matrix, $I$ is the identity matrix, $g_i^{eq}$ is the equilibrium distribution function, and $F_j$ is the force effect term. Details of $M$ and $S$ can be found in \citet{Coveney2002}. Here $g_i^{eq}$ and $F_j$ can be calculated by
\begin{eqnarray}
&&g_i^{eq}=\omega_i \rho [1+\frac{\boldsymbol{e}_i \cdot \boldsymbol{u}}{c_s^2}+ \frac{\boldsymbol{u}\boldsymbol{u}:(\boldsymbol{e}_i \boldsymbol{e}_i -c_s^2 \boldsymbol{I})}{c_s^4}],\\
&&F_j=\omega_j[\frac{\boldsymbol{e}_j-\boldsymbol{u}}{c_s^2}+\frac{\boldsymbol{e}_j \cdot \boldsymbol{u}}{c_s^4}\boldsymbol{e}_j] \cdot \boldsymbol{f},
\label{eq:Gforce}
\end{eqnarray}
where $\omega_i$ are the weights, $c_s=\Delta x /(\sqrt{3} \Delta t)$ is the the sound speed, and $\boldsymbol{f}$ is the body force. 
The kinematic viscosity $\nu$ of the fluid is determined by the relaxation time $\tau$, according to $\nu=(\tau-0.5)c_s^2\Delta t$~\citep{qian1992lattice,chen1998lattice,wang2022recent}.

The macro variables including the density, pressure and momentum can be directly calculated from the distribution functions as follows,
\begin{equation}
\rho=\sum_{i=1}^{9} g_i, \quad
p=\rho c_s^2, \quad
\rho \boldsymbol{u}=\sum_{i=1}^{9} g_i \boldsymbol{e}_i + \frac{1}{2}\boldsymbol{f}\Delta t.
\end{equation}

Here, the feedback immersed boundary method proposed~\citep[see e.g.,][]{GOLDSTEIN1993354,huang2019review} is employed to handle the no-slip boundary conditions at the fluid--structure interface, 
\begin{eqnarray}
&&\boldsymbol{f} (\boldsymbol{x}, t) = -\int_{\Gamma} \boldsymbol{F}_L (s, t) \delta_h (\boldsymbol{X}(s,t) - \boldsymbol{x}) ds,\\
&&\boldsymbol{F}_L = \alpha_{ib} \int_0^t (\boldsymbol{U}_{ib} - \boldsymbol{U}) dt + \beta_{ib} (\boldsymbol{U}_{ib} - \boldsymbol{U}),\\
&&\boldsymbol{U}_{ib} (s, t) = \int_{V} \boldsymbol{u} (x, t) \delta_h (\boldsymbol{X}(s,t) - \boldsymbol{x}) d \boldsymbol{x},
\end{eqnarray}
where $\boldsymbol{F}_L$ is the Lagrangian force, $\boldsymbol{U}_{ib}$ is the boundary velocity obtained by the interpolation from the fluid domain, $\alpha$ and $\beta$ are positive feedback constants, $\boldsymbol{u}$ is velocity of fluid nodes, $\boldsymbol{X}$ and $\boldsymbol{x}$ are respectively the coordinates of structural nodes and fluid nodes, $s$ is arc coordinate, $V$ denotes the fluid domain, $\Gamma$ denotes the structure domain, and  $\delta_h$ is the smoothed Delta function. In this work, the delta function proposed by Peskin~\citep{peskin2002immersed} is used. In addition, $\alpha_{ib} = 0$ and $\beta_{ib} = 2.0$ are used. More details can be found in Refs.~\citep{wang2018heat,tian2011efficient,xu2018novel,Huang2022pof}.

\begin{figure}
  \centerline{
  \includegraphics[width=5in]{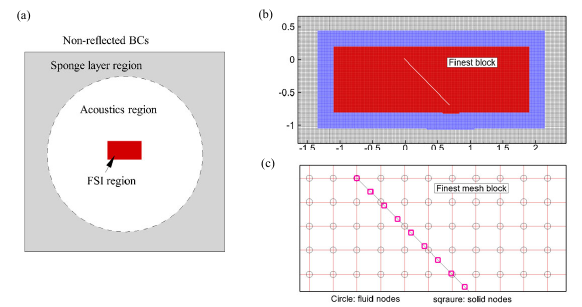}
  }
  \caption{A schematic of the multi-block for the numerical simulation: (a) the computational domain, (b) the multi-block mesh and (c) the fluid and structure nodes in the IBM.}
\label{fig:mesh}
\end{figure}

The computational domain shown in Fig.~\ref{fig:mesh}(a) extends respectively from $-150\bar{c}$ to $150\bar{c}$ in both directions for 2D and from $-50\bar{c}$ to $ 50\bar{c}$ in all three directions for 3D. A sponge layer is applied respectively at a distance of $100\bar{c}$ (2D) and $45\bar{c}$ (3D) for the outer domain to damp the solution to the far-field quiescent state, and a characteristic non-reflective boundary condition is also applied at the external boundaries to guarantee the reflections from the external boundaries are negligible. 
Seven (2D) and five (3D) levels of mesh are used in present study, and the mesh is refined with a factor of two with the finest mesh spacing of $0.005\bar{c}$ (2D) and $0.02\bar{c}$, respectively. Three mesh blocks close to the wing is shown in Fig.~\ref{fig:mesh}(b) to illustrate the mesh strategy. The finest mesh covers the sweeping range of the wing for the near-field flow, and the acoustics are far away from the wing is resolved with mesh resolution of $0.16\bar{c}$ namely the acoustics region in Fig.~\ref{fig:mesh}(a). The Lagrangian mesh spacing is about $0.5\Delta x$, as demonstrated in Fig.~\ref{fig:mesh}(c).
In the current simulation, a reference density $\rho=1.0$ and a sound velocity $c_s=1/\sqrt{3}$ are used. The sound pressure is nondimensionalised by $\rho c_s^2$. The validation of the numerical solver is given in Appendix~\ref{appA}, and the mesh convergence study is provided in Appendix~\ref{appB}. It should be noted that the force and sound power are transformed into the physical space assuming a wing mean chord length of $2 mm$ in earth atmosphere environment, to make it intuitive for readers.
 
\section{Results and discussions}\label{sec:resdic}

\subsection{Purely translating wings}\label{sec:transwing}
The purely translating wing is first considered here for its simplicity. For a steady wing at a given angle of attack (see Fig.~\ref{fig:sch} a), the flow separates near the leading edge of the wing and forms a leading edge vortex (LEV). The strength of the LEV is related to the shear flow on the wing, and the aerodynamic force generated by the wing can be estimated from the strength of the LEV by using the potential flow theory as
$F = \rho \Gamma \bar{c}$ with $\Gamma$ being the strength of the circulation~\citep{dickinson1996unsteady,wu2007vorticity}. For a translating wing, the aerodynamic force can be estimated alternatively as~\citep{lee2015scaling}
\begin{equation}
F = \Delta I / T,
\label{eq:fortrans}
\end{equation}
where $\Delta I$ is the momentum induced by the vortical structures and $T$ is the translating period of the wing. $\Delta I \sim \rho \Gamma S$ and $S = \bar{c}$ in 2D. Therefore, we obtain $F \sim \rho \Gamma  f \bar{c}$ by using Eq.~\ref{eq:fortrans}. According to the potential flow theory, the circulation can be expressed as $\Gamma \sim U \bar{c} \cos(\alpha_m)$ for a wing translating a given angle of attack, where $U$ is the translating velocity of the wing and $\alpha_m$ is the angle of attack which is a constant here. Therefore, the aerodynamic force generated by the translating wing can be written as
\begin{equation}
     F \sim \rho U f \bar{c}^2 \cos({\alpha_m}),
\end{equation}
and the lift ($L$) is the projection of $F$ in the vertical direction, i.e., 
\begin{equation}
    L \sim \rho U  f \bar{c}^2 \sin({2\alpha_m}).
\end{equation}

For the purely translating wing, the maximum velocity of the wing $U_m \sim A_m f$ according to the kinematics described in Eq.~\ref{eq:trankint}. Therefore, the amplitude of the lift follows
\begin{equation}
    L_m \sim \rho A_m f^2 \bar{c}^2  \sin({2\alpha_m}).
\label{eq:scallifttrans}
\end{equation}
It should be noted that $L_m$ is the amplitude of the lift rather than the mean lift. The mean lift is expected to be zero for the translating wing due to the symmetric lift during left and right translation stages.

According to the FW-H equation, the far-field sound generated by a moving objects has three components with its solution consisting of a surface integral of the monopole source, a surface integral of the dipole source and a volume integral of the quadrupole source. If the sources are compact and the observation location is far away from the wing, the solution of the  FW-H equation becomes~\citep{Tian2020cm,Seo2020bb},
\begin{equation}
    \Delta p \sim \frac{d\boldsymbol{F}}{dt}\cdot \hat{\boldsymbol{r}},
    \label{eq:dldt}
\end{equation}
where $\boldsymbol{F}$ is the aerodynamic force which is mainly determined by the pressure, and $\hat{\boldsymbol{r}}$ is the unit vector pointing from the wing to the far-field observer. Using the scaling law for the amplitude of lift in Eq.~\ref{eq:scallifttrans}, the scaling law for the amplitude of sound pressure generated in the vertical direction is
\begin{equation}
    \Delta p \sim L_m f \sim \rho A_m  f^3 \bar{c}^2 \sin({2\alpha_m}).
    \label{eq:scaldptrans}
\end{equation}

To confirm the validity of the constructed scaling laws for the amplitude of lift and sound generated in the vertical direction, direct numerical simulation is conducted by using an IB-LBM method as mentioned in \S~\ref{sec:nummthd} with the parameters shown in Tab.~\ref{tab:para} being considered which gives 35 simulations in total. It should be noted that we randomly chose parameters shown in Tab.~~\ref{tab:para} instead of simulating all parameters to save computational resource and cover the listed parameter range. 

\begin{figure}
  \centerline{\includegraphics[width=5.8in]{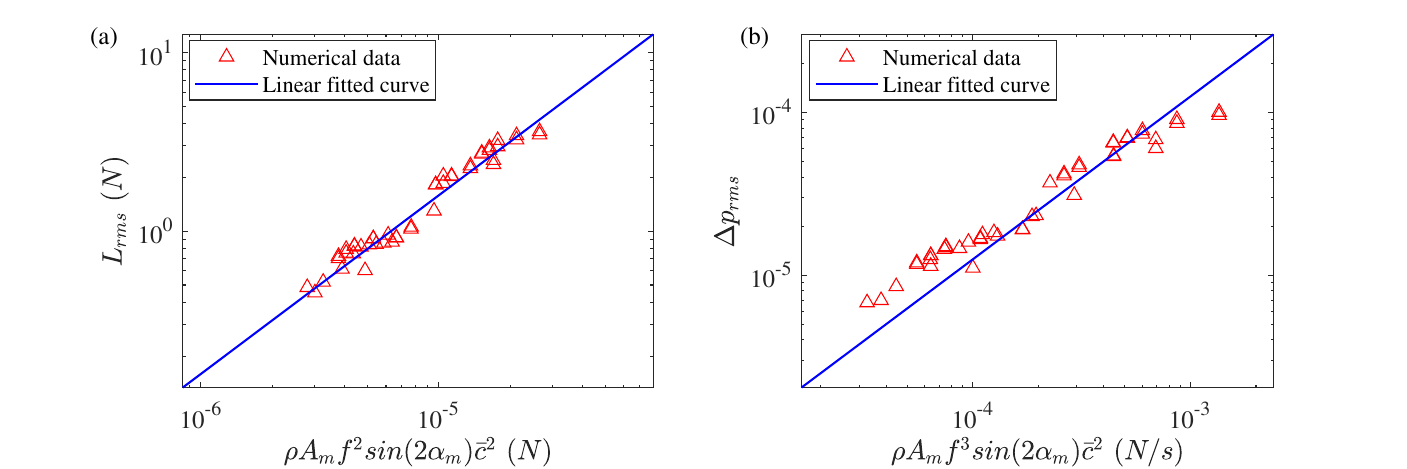}}
  \caption{Scaling laws of a translating wing: (a) the lift, and (b) the sound pressure $\theta = 90^o$ and $r=85\bar{c}$ scaled by using the fluid density and sound speed at quiescent state.}
\label{fig:trans}
\end{figure}

Fig.~\ref{fig:trans} shows the amplitudes of lift and the root-mean-square of sound pressure in the vertical direction generated by the translating wing from 40 numerical simulations according to scaling laws in  Eqs.~\ref{eq:scallifttrans} and \ref{eq:scaldptrans}. It shows that the lift collapses to a straight line confirming the validity of the derived scaling laws. The sound pressure also collapses to a line as expected in the scaling law, although it shows slightly larger discrepancies compared with the lift which could be attributed to the non-perfect sinusoidal sound pressure in time domain due to the viscous effects.

\begin{figure}
  \centerline{\includegraphics[width=5.8in]{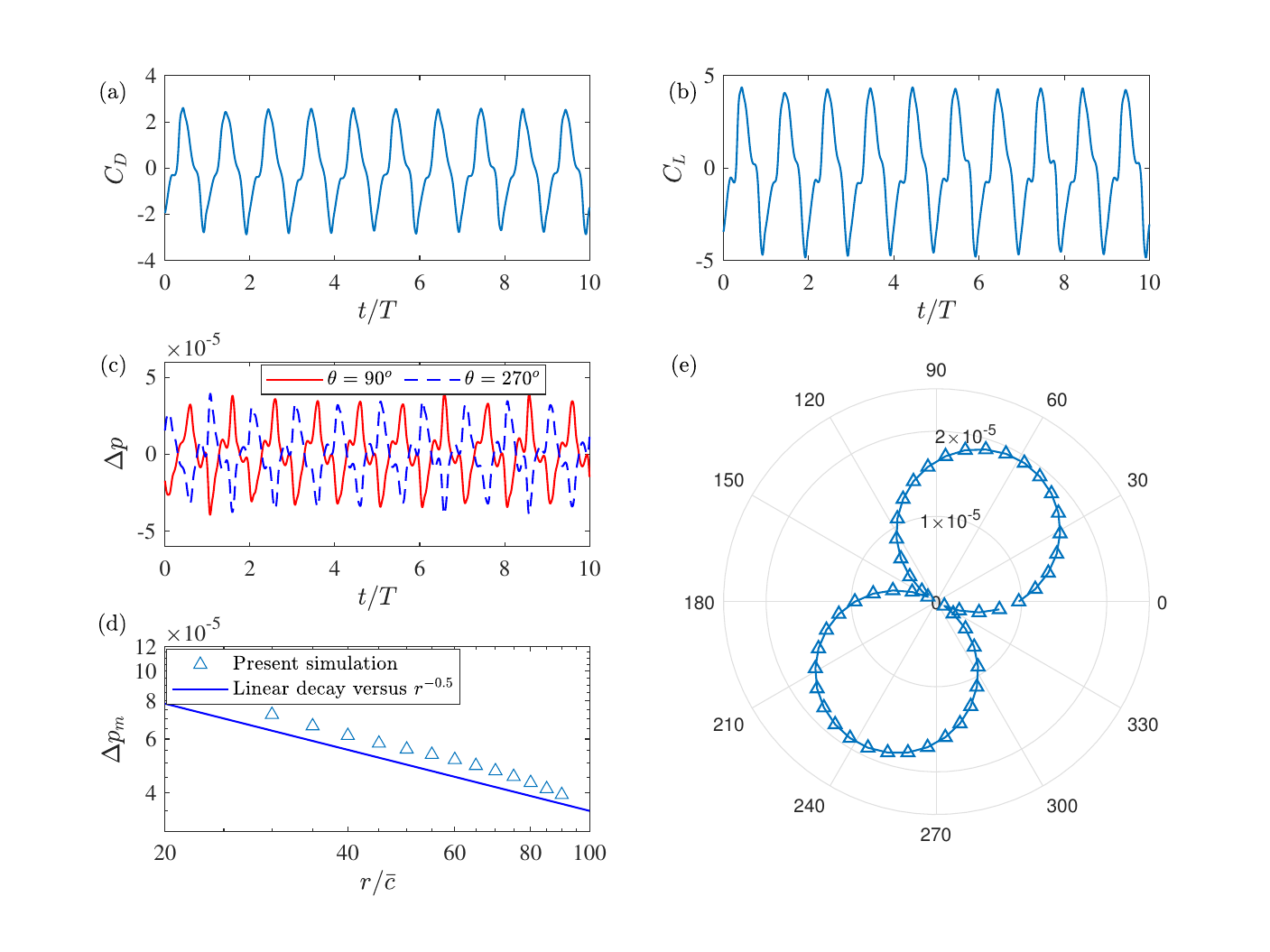}}
  \caption{Sound generation by a translating wing at $M=0.05$, $Re=400$, $\alpha_m=30^o$ and $A_m=\bar{c}$: (a) time histories of $C_D$ (horizontal direction); (a) time histories of $C_L$ (vertical direction); (c) time histories of sound pressure measured at a distance of $100\bar{c}$, $\theta=90^o$ and $270^o$; (d) decay of sound pressure with distance measured at $\theta=90^o$; (e) sound pressure directivity measured at a distance of $100\bar{c}$.}
\label{fig:transdp}
\end{figure}

\begin{figure}
  \centerline{
  \includegraphics[width=5in]{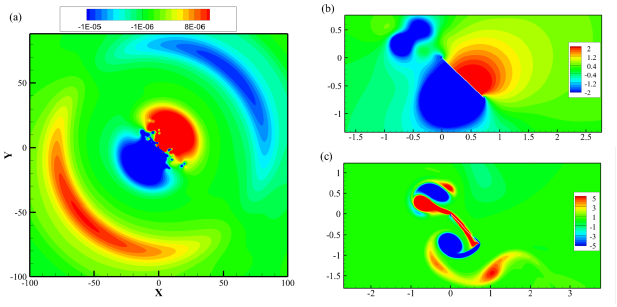}
  }
  \caption{Instantaneous acoustics and flow fields generated by a translating wing at $t=0T/4$ with $M=0.04$, $Re=400$, $\alpha_m=45^o$ and $A_m=0.8 \bar{c}$: (a) sound pressure contours; (b) pressure coefficient contours near the wing; (c) vorticity contours near the wing.}
\label{fig:transcontour}
\end{figure}

To illustrate the major features of both force and sound generations by a translating wing, Fig.~\ref{fig:transdp} shows the force histories, sound pressure time histories, decay of sound pressure with distance and sound pressure directivity; and Fig.~\ref{fig:transcontour} shows the instantaneous acoustics and flow fields at $t=0T/4$ with $M=0.04$, $Re=400$, $\alpha_m=45^o$ and $A_m=0.8 \bar{c}$. Both drag and lift forces are periodic and symmetric (see Fig.~\ref{fig:transdp} (a) and (b)). This is because that when the wing translates from the left to the right, a suction region and two vortices are generated at bottom side of the wing as shown in Fig.~\ref{fig:transcontour} (b) and (c), which are ``rotational symmetric" to those generated when the wing moves from the right to the left (not shown for conciseness). Due to the compact sound sources,  the sound field is quasi-symmetric about the wing, as shown in Fig.~\ref{fig:transdp} (a). A quantitative illustration is shown in Fig.~\ref{fig:transdp} (c), which measures the sound pressure at a distance of $100 \bar{c}$ and $\theta = 90^o$ and $270^o$. It shows that the measured sound pressure signals are close to a sinusoidal shape which aligns with the prescribed kinematics, and the same amplitude and out-of-phase pressure at the two measure points confirms the symmetric sound fields, as shown in the acoustics field in Fig.~\ref{fig:transcontour} (a) and the polar plot of the sound diretivity in Fig.~\ref{fig:transdp} (e). In addition, the decay of sound pressure measured at $\theta=90^o$ with the distance is shown in Fig.~\ref{fig:transdp} (d), which agrees well with the linear relation from theoretical analysis. 

\subsection{Translating and pitching wings}{\label{sec:trph}}

\begin{figure}
  \centerline{\includegraphics[width=2.5in]{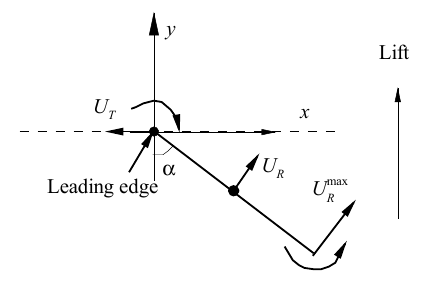}}
  \caption{A schematic of the velocity of a translating and pitching wing.}
\label{fig:transpitch}
\end{figure}

In the real world, insects and birds normally flap theirs wings with a pitching motion to get high aerodynamic performance. Here, the wing with such a combined flapping and pitching motion is further considered. Fig.~\ref{fig:transpitch} shows a schematic of the wing velocity when the combined kinematics is used. In addition to the circulation generated due to the leading edge vortex induced by the translation, the pitching motion induce extra circulation which could enhance the lift generation and should be considered. According to the potential flow theory~\citep[see e.g.,][]{dickinson1996unsteady,wu2007vorticity}, the force is determined by
\begin{equation}
    F = F_T +F_R \sim \rho (\Gamma_T + \Gamma_R) f \bar{c},
\end{equation}
where $F_T$ and $F_R$ are respectively the force caused by the translating and pitching components, and $\Gamma_T$ and $\Gamma_R$ are the circulation induced by the translation and pitching, respectively. According to Section~\ref{sec:transwing}, $F_T \sim \rho \Gamma_T f \bar{c} \sim \rho U_T  f \bar{c}^2 \sin{(0.5\pi-\alpha_m)} = \rho A_m f^2 \bar{c}^2  \sin(0.5\pi-\alpha_m)$. The force generated by the pitching can be written as $F_R \sim \rho \Gamma_R f \bar{c} \sim \rho U_R f  \bar{c}^2$, where $U_R$ is the averaged rotating velocity along the chord of the wing, i.e., the rotating velocity measured at the middle of the wing chord. According to the kinematics described in Eq.~\ref{eq:pitckint}, $U_R \sim 0.5 \alpha_m f \bar{c}$, and thus the amplitude of the aerodynamic force generated by a wing with combined translation and pitching motion follows
\begin{equation}
    F_m \sim \rho A_m f^2  \bar{c}^2\sin(0.5 \pi-\alpha_m) + 0.5 \rho \alpha_m  f^2 \bar{c}^3.
\end{equation}
By projecting the force on to the vertical direction, then the amplitude of the lift has the following scaling law
\begin{equation}
    L_m \sim \lambda \rho A_m f^2  \bar{c}^2  \sin(2 \alpha_m), 
    \label{eq:scallifttrans&pitch}
\end{equation}
with 
\begin{equation}
    \lambda = 1+\frac{\alpha_m \bar{c}}{2A_m \sin \alpha_m}. 
    \label{eq:lamda}
\end{equation} 
By using Eq.~\ref{eq:dldt}, a scaling law for the amplitude of the sound generated by a wing with combined translation and pitching motion in the vertical direction is written as
\begin{equation}
    \Delta p \sim \frac{dL}{dt} \sim \lambda \rho A_m  f^3 \bar{c}^2 \sin(2\alpha_m).
    \label{eq:scaldptptra&ptch}
\end{equation}


\begin{figure}
  \centerline{\includegraphics[width=4.in]{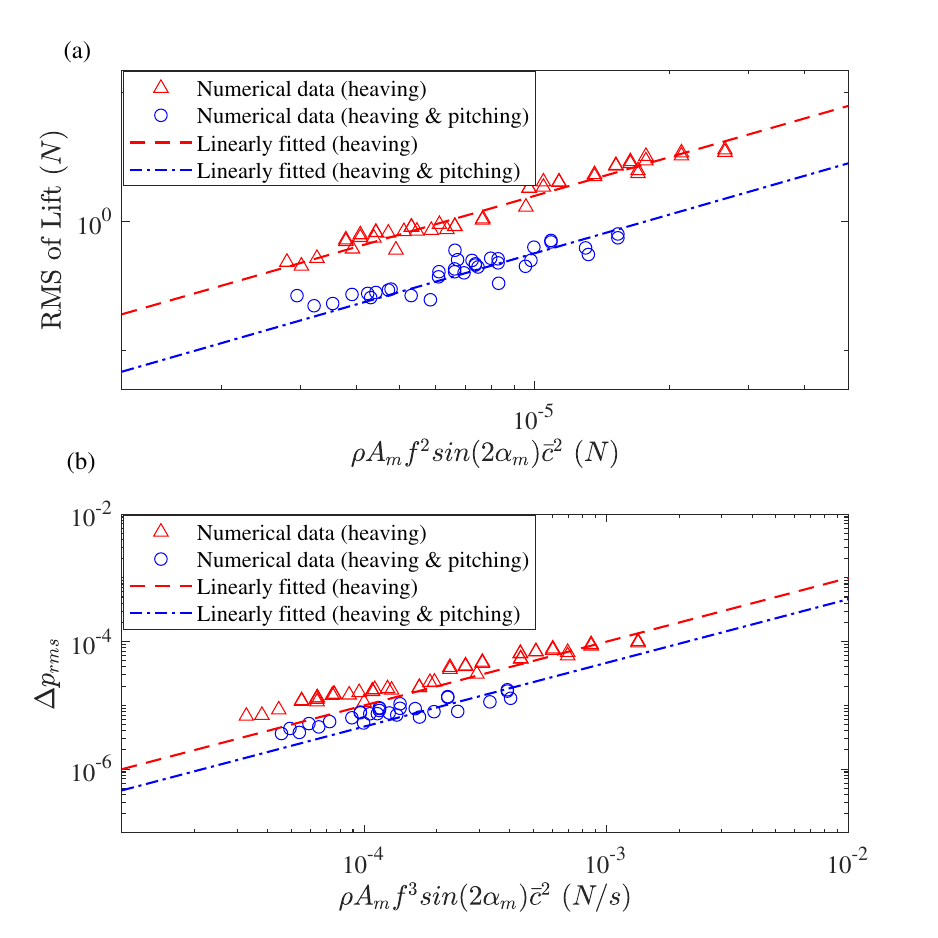}}
  \caption{Scaling laws of 2D wings undergoing translating only, combined translating and pitching motion: (a) the amplitude of lift, and (b) sound pressure at $\theta = 90^o$ and $r=85 \bar{c}$ scaled by using the fluid density and sound speed at quiescent state. $\lambda$ is unit for the translating wing and determined by Eq.~\ref{eq:lamda} for the translating and pitching wing.}
\label{fig:scalp2dall}
\end{figure}

It is noted that the scaling laws described by Eqs.~\ref{eq:scallifttrans&pitch} and~\ref{eq:scaldptptra&ptch} for the wing undergoing combined translating and pitching motion are consistent with those described by Eqs.~\ref{eq:scallifttrans} and \ref{eq:scaldptrans} for the wing undergoing purely translating motion. The coefficient $\lambda$ is derived for the modification due to the pitching motion. To confirm the validity of the scaling laws for a wing with combined translation and pitching kinematics, another 30 simulations are conducted with the parameters selected from the Tab.~\ref{tab:para}. Fig.~\ref{fig:scalp2dall} shows the lift and sound pressure from all numerical simulations. The results show that the numerical data collapse well to a line for both the lift and sound pressure. It is also noted that the wings with pitching motion generate smaller lift amplitudes and thus smaller acoustic outputs compared with those wings translating only, as the pitching motion considered in this work relieves the relative velocity near the wing tip during the start of each stroke to the flow as discussed in~\cite{wang2019acoustics}. 


\subsection{Three-dimensional flapping wings}{\label{sec:hoving3d}}

In this section, we further examine three-dimensional flapping wings in hovering flight. According to the potential flow theory discussed in Section.~\ref{sec:transwing}, the aerodynamic force generated by the flapping wing $F \sim \rho \Gamma S f$, where the area $S = AR \bar{c}^2$, circulation $\Gamma \sim U \bar{c} \sin(0.5 \pi - \alpha_m)$ and the maximum wing tip velocity $U_m \sim \phi_m f AR \bar{c}$. Take the aspect ratio effects $\frac{AR}{AR + 2}$ into account, a scaling law of the amplitude of lift and sound pressure in the vertical direction for the 3D flapping wing can be obtained as
\begin{eqnarray}
    L_m = F \cos(0.5 \pi - \alpha_m) \sim \rho \lambda \phi_m f^2 \sin(2 \alpha_m) AR^2 \bar{c}^4 \frac{AR}{AR + 2},\\
    \Delta p  \sim \frac{dL}{dt} \sim \lambda \rho \phi_m f^3 AR^2 \bar{c}^4 \sin(2\alpha_m) \frac{AR}{AR + 2},
\end{eqnarray}
where $\lambda$ is the same as that in Eq.~\ref{eq:lamda}.

\begin{figure}
  \centerline{\includegraphics[width=5.8in]{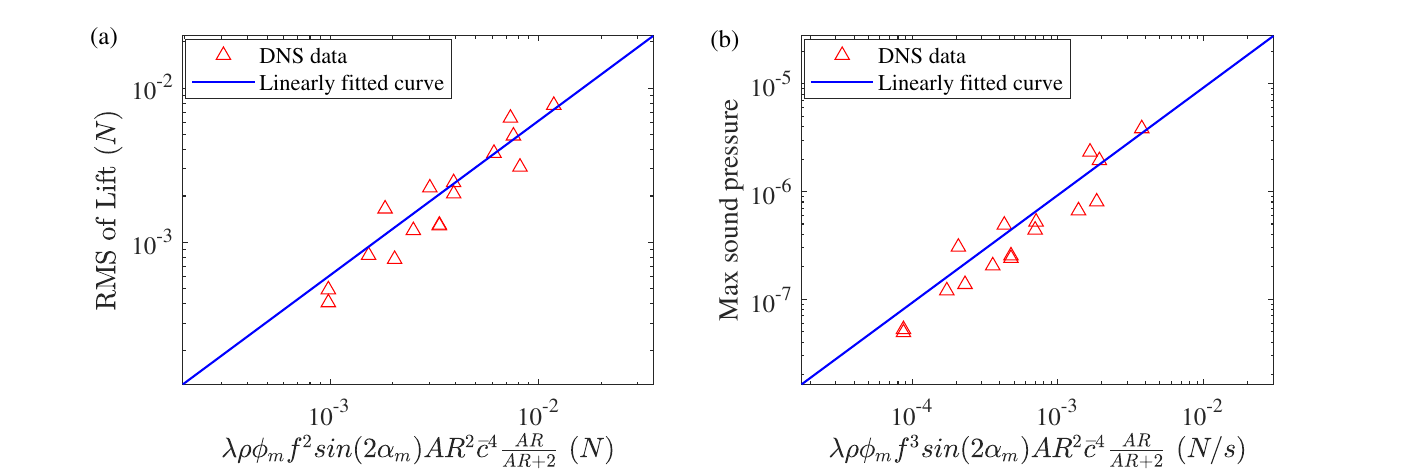}}
  \caption{Scaling laws of a 3D wing undergoing combined stroke and pitching motion: (a) the lift, and (b) sound pressure in the vertical direction at $r=40\bar{c}$ scaled by using the fluid density and sound speed at quiescent state.}
\label{fig:scal3damp}
\end{figure}

\begin{figure}
  \centerline{\includegraphics[width=5.8in]{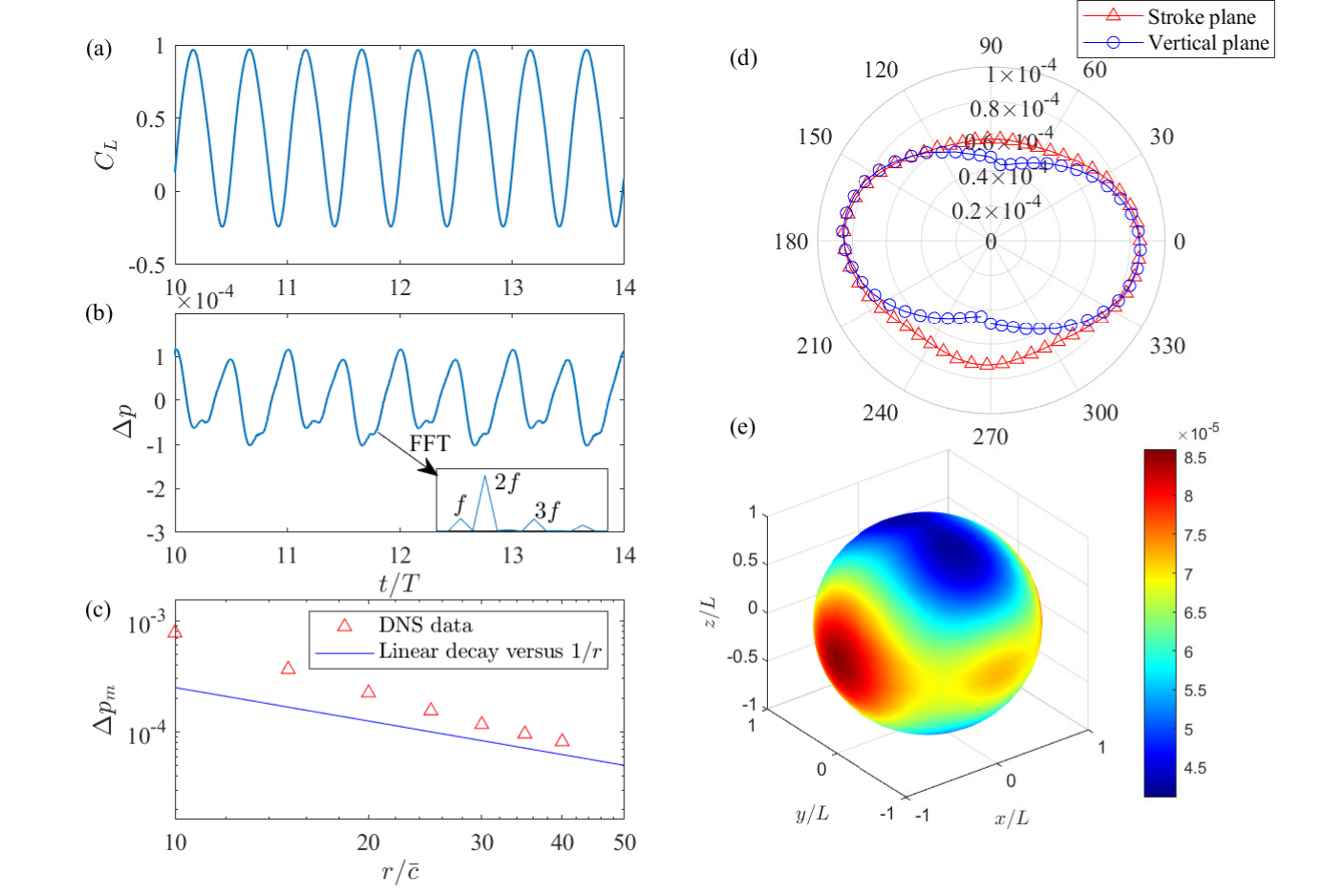}}
  \caption{Sound generation by a 3D flapping wing at $M=0.05$, $Re=200$, $\phi_m=45^o$, $\alpha_m=45^o$ and $AR=2.0$: (a) time histories of $C_L$; (b) time histories of sound pressure measured at a distance of $30\bar{c}$ on the top of the wing; (c) decay of sound pressure with distance measured on the top of the wing; (d) sound pressure directivity measured at a distance of $40\bar{c}$ on the stroke plane and vertical plane; (e) sound pressure directivity measured over a spherical surface of a radius of $40\bar{c}$.}
\label{fig:his3d}
\end{figure}

 To confirm the validity of the scaling laws for the 3D flapping wing in hovering flight, 16 direct numerical simulations are conducted with the parameters selected from Tab.~\ref{tab:para} covering a range of Reynolds numbers, Mach numbers, wing kinematics and aspect ratios. According to \citet{ellington1996leading}, the insect wing can be characterised by $\hat{r}$ which is normally in a range between 0.4 and 0.6. Here $\hat{r} = 0.53$ in the middle range is selected to reduce the parameters in the simulation. Fig.~\ref{fig:scal3damp} shows the constructed scaling laws with the data from DNS which shows that the numerical results are well positioned around the scaling laws confirming the validity of the scaling laws for the aerodynamic force and the sound pressure.
 
 To demonstrate the properties of aerodynamic force and sound generated by 3D flapping wing, a specific case at $M=0.05$, $Re=200$, $\phi_m=45^o$, $\alpha_m=45^o$ and $AR=2.0$ are selected for analysis, as shown in Fig.~\ref{fig:his3d}. It is found that the lift coefficient is quasi-sinusoidal (see Fig.~\ref{fig:his3d}a), while the sound pressure measured at a distance of $40 \bar{c}$ on the top of the wing reaches a periodic but non-sinusoidal state (see Fig.~\ref{fig:his3d}b). Although complex acoustic features in time space are observed in Fig.~~\ref{fig:his3d} (b), the second harmonic is the dominating frequency of the sound, and is consistent with the frequency of the lift force (as shown by the insert which is the Fast Fourier Transform of the sound pressure). This observation agrees well with that observed by~\citet{sueur2005sound} considering real flies with more complex wing kinematics. The coincidence of the frequencies of the lift and the sound pressure explains why a simple scaling law can be constructed by relating the aeroacoutics to the aerodynamic force directly without losing the major physics, as shown in Fig.~\ref{fig:scal3damp} (a) and (b). Fig.~\ref{fig:his3d} (c) shows the decay of the pressure measured on the top of the wing versus the distance, which agrees well with the linear decay law in the far-field region, as indicated by the solid line confirming the accuracy of the numerical simulation. For the near field with $r/\bar{c} < 25$, it also shows that the decay of sound pressure does not follow the linear trend due to the near-field effects.  
 This is the reason that the sound pressure at a distance of $40\bar{c}$ is used  in Fig.~\ref{fig:scal3damp} to validate the scaling law eliminating the effects from the near-field unsteady flow. Fig.~\ref{fig:his3d} (d) shows the directivity of the sound pressure on the stroke plane ($xy$) and vertical plane ($yz$), indicating a nearly symmetric profile about the $90^\circ$ line due to the symmetric kinematics. A more comprehensive directivity in 3D space is shown in Fig.~\ref{fig:his3d} (e), which shows that the maximum sound is observed on the stroke plane, indicating that the scaling law for the maximum sound pressure in the vertical direction corresponding to the lift does not necessarily reflect the overall acoustic outputs.

\begin{figure}
  \centerline{\includegraphics[width=5.8in]{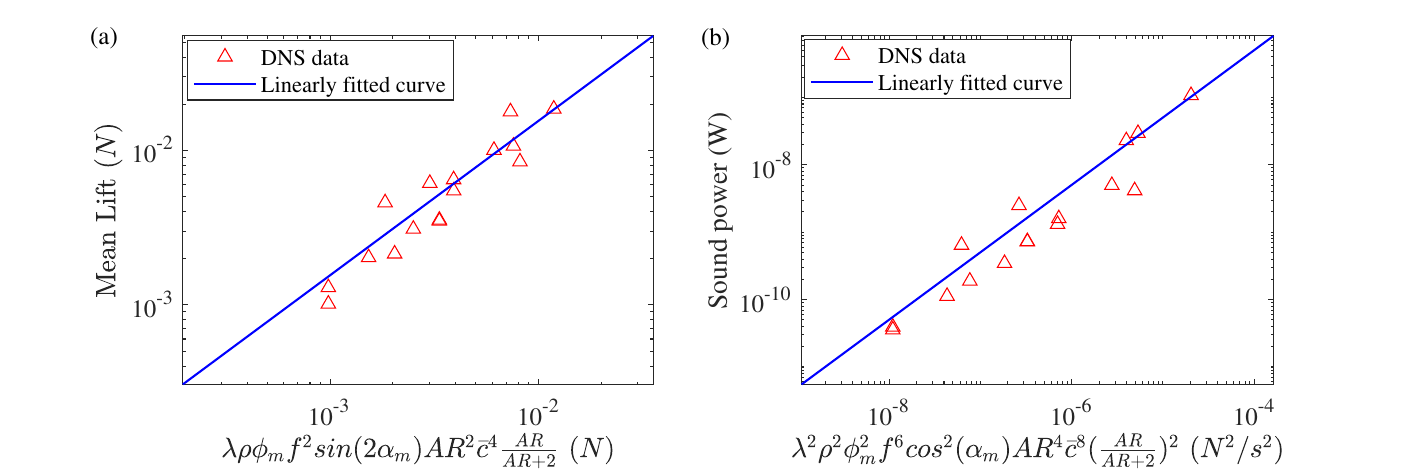}}
  \caption{Scaling laws for the mean lift (a), and sound power at $r=40\bar{c}$ (b) of the 3D hovering flapping wing.}
\label{fig:scal3d}
\end{figure}

 The mean lift and the sound power are also important in measuring the flapping performance. For the hovering flapping wing considered in this work, the mean lift is proportional to its amplitude, as only small negative lift is generated as shown in Fig.~\ref{fig:his3d} (a). This limited negative lift is actually common in hovering flight as observed by~\citet{nagai2010experimental} in experiment and~\citet{young2009details} in simulation with a pair of flexible wings. Therefore, the mean lift ($\bar{L}$) should follow the same scaling law as the amplitude of the lift, which can be given as
\begin{eqnarray}
    \bar{L} &\sim& \lambda \rho \phi_m f^2 \sin(2 \alpha_m) AR^2 \bar{c}^4 \frac{AR}{AR + 2}.
    \label{eq:scalmeanL}
\end{eqnarray}
It should be noted that this scaling law is consistent with that obtained by~\citet{lee2015scaling}, who ignored the effects from pitching (as reflected by $\lambda$) so that their scaling law covers a large range of lift (up to 2nd-order difference) where the pitching effects are not significant. The consistent scaling law is constructed based on the fact that the mean lift scales with the lift amplitude in hovering flight which was not explained in \citet{lee2015scaling}. The current DNS data shown in Fig.~\ref{fig:scal3d} (a) also confirms the validity of the scaling law in Eq.~\ref{eq:scalmeanL}. The sound power ($SP$) which measures the overall acoustic outputs is further obtained as  
 \begin{eqnarray}
    SP &\sim& (\Delta p)^2  \sim \left(\frac{dF}{dt}\right)^2\sim \lambda^2 \rho\phi_m^2 f^6 AR^4 \bar{c}^8 \cos^2 (\alpha_m) (\frac{AR}{AR + 2})^2,
\end{eqnarray}
 which is confirmed to capture the major physics well by the data from DNS, as shown in Fig.~\ref{fig:scal3d}(b). It is noted that the derived scaling law is consistent with Curl's theory indicating that the dipole sound power is proportional to the sixth power of the flow velocity~\citep{curle1955influence}. This also indicates that the sound generated by the flapping wing is dominated by the dipole sound and/or the combination of multiple dipole sounds, as shown in Fig.~\ref{fig:his3d} (d) and (e). 

\section{Conclusions}\label{sec:conc}
This work presents the scaling laws for the sound generation of bio-inspired flapping wings during hovering flight, which are derived by using the classical quasi-steady flow theory and the FW-H acoustic equation. Three groups of wings are consider: 2D purely translating wings, 2D translating and pitching wings, and 3D stroking and pitching wings. Direct numerical simulations considering a range of parameters including the Reynolds number, Mach number and wing kinematics confirm that the constructed scaling laws capture the major physics and agree well with the direct numerical simulation results. 

It is noted that the simple scaling laws directly relate the kinematics of bio-inspired flapping wings to their noise emission. The scaling laws are able to explain the properties of sound generated by most insects with different wing sizes and kinematics. The success of this work is due to the fact that the major sound by the bio-inspired flapping wings considered here is the Gutin sound. The scaling analysis of other sound sources should be considered for complicated wings and high-speed flight, which will be our future effort.  



\section*{Funding}\label{sec:ack}
This work was supported by the Australian Research Council Discovery Project (grant No. DP200101500) and conducted with the assistance of computational resources from the National Computational Infrastructure (NCI), which is supported by the Australian Government.

\section*{Declaration of interests}
The authors report no conflict of interest.


\appendix
\section{Validation of the numerical solver}\label{appA}
This section presents a validation of the numerical solver considering the aerodynamics and acoustics of flow around a cylinder at $Re=\rho UD / \mu = 150$ and $M=U/c_s=0.2$ (where $\rho$ and $\mu$ are respectively the density and viscosity of the fluid, $D$ is the diameter of the cylinder, $U$ is the velocity of the uniform incoming flow and $c_s$ is the sound speed of the fluid flow at quiescent state), as which is a benchmark and has been extensively validated by~\citet{inoue2002sound,wang2020immersed}. 

\begin{figure}
  \centerline{\includegraphics[width=4.in]{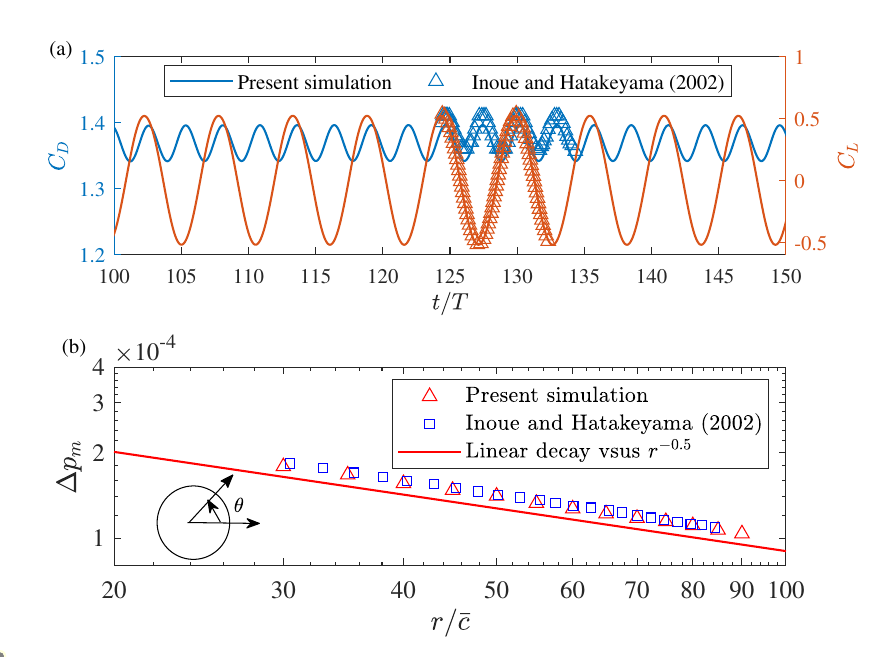}}
  \caption{Flow around a circular cylinder at $Re=150$ and $M=0.2$: (a) the time histories of $C_L$ and $C_D$; (b) the decay of sound pressure measured at $\theta=90^o$ versus distance.}
\label{fig:cy}
\end{figure}

In this validation, the fluid domain has a dimension of $150D \times 150D$, and six mesh blocks are used with the finest mesh spacing of $D/100$. The quantitative comparison of the lift and drag coefficients, as well as the sound pressure measured at the top of the cylinder are shown in Fig.~\ref{fig:cy}. As we can see, the lift and drag coefficients predicted by the current method agree well with the data from \citet{inoue2002sound}. In addition, the sound pressure with the distance from the cylinder centre is proportional to $1/sqrt{r}$, which is consistent with the result reported in the reference. Therefore, the IB-LBM solver for the direct simulation of acoustics at low Reynolds numbers is reliable.

\section{Mesh convergence study}\label{appB}

\begin{figure}
  \centerline{\includegraphics[width=5.8in]{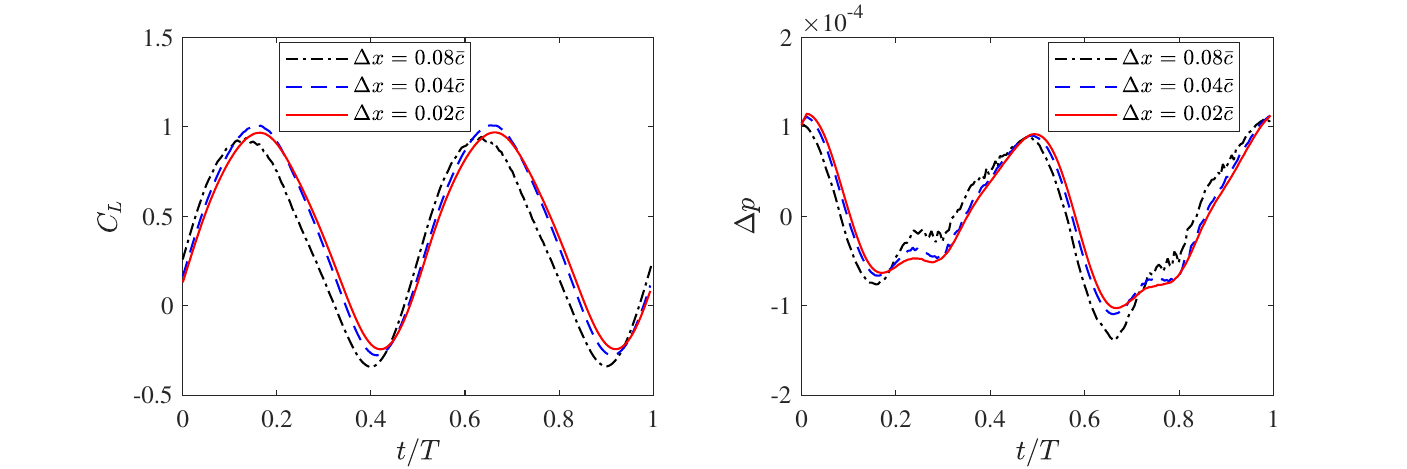}}
  \caption{Mesh convergence study of 3D flapping wing at $M=0.05$, $Re=200$, $\phi_m=45^o$, $\alpha_m=45^o$ and $AR=2.0$: (a) time histories of $C_L$, and (b) time histories of sound pressure measured above the wing at a distance of $30\bar{c}$.}
\label{fig:meshconvg3d}
\end{figure}

This section presents a mesh convergence study of the numerical solver considering the aerodynamics and acoustics of a 3D flapping wing in hovering flight. All the parameters considered here are same as those used in Fig.~\ref{fig:his3d}. Three mesh sizes in the near-field region are considered, i.e., $\Delta x = 0.08 \bar{c}, 0.04 \bar{c}$ and $0.02\bar{c}$. Fig.~\ref{fig:meshconvg3d} shows the time histories of lift coefficient $C_L$ and the sound pressure $\Delta p$ measured above the wing at a distance of $30\bar{c}$. Both of  $C_L$ and $\Delta p$ show a good convergence with the refinement of the mesh. The finest mesh of $0.02\bar{c}$ is used for the current study to provide a better sound visualisation.

\bibliographystyle{jfm}
\bibliography{jfm-instructions}

\end{document}